\documentclass[letterpaper]{article} 
\usepackage{aaai24}  
\usepackage{times}  
\usepackage{helvet}  
\usepackage{courier}  
\usepackage[hyphens]{url}  
\usepackage{graphicx} 
\usepackage{natbib}  
\usepackage{caption} 
\usepackage{algorithm}
\usepackage{algorithmic}

\usepackage{stmaryrd}
\usepackage{amsmath}
\usepackage{amssymb}
\usepackage{mathtools}
\usepackage{amsthm}
\usepackage{dsfont}
\usepackage{hyperref}
\hypersetup{
	colorlinks=true,
	linkcolor=red,
	filecolor=blue,      
	urlcolor=blue,
	citecolor=violet,
}
\usepackage{diagbox}
\usepackage{adjustbox}
\usepackage{multirow}
\usepackage{makecell}
\usepackage{subfigure}
\usepackage{graphicx}

\usepackage{caption}

\theoremstyle{plain}

\newtheorem{proposition}{Proposition}

\theoremstyle{definition}

\theoremstyle{remark}

\usepackage{xspace}
\usepackage[capitalize,noabbrev]{cleveref}
\usepackage[textsize=tiny]{todonotes}
\crefname{section}{Sec.}{Secs.}
\Crefname{section}{Section}{Sections}
\Crefname{table}{Table}{Tables}
\crefname{table}{Tab.}{Tabs.}
\usepackage{yhmath}
\usepackage{pifont}
\newcommand{\method}{\textsc{BindDM}\xspace}

\usepackage{newfloat}
\usepackage{listings}
\DeclareCaptionStyle{ruled}{labelfont=normalfont,labelsep=colon,strut=off} 
\floatstyle{ruled}
\newfloat{listing}{tb}{lst}{}
\floatname{listing}{Listing}
\title{Binding-Adaptive Diffusion Models for Structure-Based Drug Design}
\author{
    Zhilin Huang\textsuperscript{\rm 1,2}\equalcontrib,
    Ling Yang\textsuperscript{\rm 3}\equalcontrib,
    Zaixi Zhang\textsuperscript{\rm 4},
    Xiangxin Zhou\textsuperscript{\rm 5},\\
    Yu Bao\textsuperscript{\rm 6},
    Xiawu Zheng\textsuperscript{\rm 2},
    Yuwei Yang\textsuperscript{\rm 6},
    Yu Wang\textsuperscript{\rm 2\dag}, 
    Wenming Yang\textsuperscript{\rm 1,2}\thanks{Corresponding author.}\\
    \url{https://github.com/YangLing0818/BindDM}
}
\affiliations{
    \textsuperscript{\rm 1}Shenzhen International Graduate School, Tsinghua University,
    \textsuperscript{\rm 2}Peng Cheng Laboratory,
    \textsuperscript{\rm 3}Peking University,\\
    \textsuperscript{\rm 4}University of Science and Technology of China,
    \textsuperscript{\rm 5}University of Chinese Academy of Sciences,
    \textsuperscript{\rm 6}ByteDance,
    zerinhwang03@pku.edu.cn, zaixi@mail.ustc.edu.cn, \{zhouxiangxin1998, nlp.baoy\}@gmail.com, yuwei.yang@bytedance.com, \{zhengxw01, wangy20\}@pcl.ac.cn, \{yangling0818, yangelwm\}@163.com
}

\usepackage{bibentry}

\usepackage{amsmath,amsfonts,bm}

\def\eqref#1{equation~\ref{#1}}

\def\1{\bm{1}}

\def\rvc{{\mathbf{c}}}

\def\rve{{\mathbf{e}}}

\def\rvh{{\mathbf{h}}}

\def\rvs{{\mathbf{s}}}

\def\rvv{{\mathbf{v}}}

\def\rvx{{\mathbf{x}}}

\def\rmA{{\mathbf{A}}}

\def\rmC{{\mathbf{C}}}
\def\rmD{{\mathbf{D}}}

\def\rmH{{\mathbf{H}}}

\def\rmM{{\mathbf{M}}}

\def\rmP{{\mathbf{P}}}

\def\rmV{{\mathbf{V}}}

\def\rmX{{\mathbf{X}}}

\def\rmZ{{\mathbf{Z}}}

\def\vmu{{\bm{\mu}}}

\def\vc{{\bm{c}}}

\def\vv{{\bm{v}}}

\def\vx{{\bm{x}}}

\def\mI{{\bm{I}}}

\DeclareMathAlphabet{\mathsfit}{\encodingdefault}{\sfdefault}{m}{sl}
\SetMathAlphabet{\mathsfit}{bold}{\encodingdefault}{\sfdefault}{bx}{n}

\def\gM{{\mathcal{M}}}

\def\gP{{\mathcal{P}}}

\def\gU{{\mathcal{U}}}

\newcommand{\R}{\mathbb{R}}

\begin{document}

\maketitle
\begin{abstract}
Structure-based drug design (SBDD) aims to generate 3D ligand molecules that bind to specific protein targets.
Existing 3D deep generative models including diffusion models have shown great promise for SBDD.
However, it is complex to capture the essential protein-ligand interactions exactly in 3D space for molecular generation.
To address this problem, we propose a novel framework, namely \textbf{Bind}ing-Adaptive \textbf{D}iffusion \textbf{M}odels (\method).
In \method, we adaptively extract \textit{subcomplex}, the essential part of binding sites responsible for protein-ligand interactions.
Then the selected protein-ligand subcomplex is processed with SE(3)-equivariant neural networks, and transmitted back to each atom of the complex for augmenting the target-aware 3D molecule diffusion generation with binding interaction information. We iterate this hierarchical complex-subcomplex process with \textit{cross-hierarchy interaction node} for adequately fusing global binding context between the complex and its corresponding subcomplex. Empirical studies on the CrossDocked2020 dataset show \method can generate molecules with more realistic 3D structures and higher binding affinities towards the protein targets, with up to \textbf{-5.92} Avg. Vina Score, while maintaining proper molecular properties. Our code is available at \href{https://github.com/YangLing0818/BindDM}{https://github.com/YangLing0818/BindDM}
\end{abstract}

\section{Introduction}
Designing ligand molecules that can bind to specific protein targets and modulate their function, also known as \textit{structure-based drug design} (SBDD)~\cite{anderson2003process,batool2019structure}, is a fundamental problem in drug discovery and can lead to significant therapeutic benefits. 
SBDD requires models to synthesize drug-like molecules with stable 3D structures and high
binding affinities to the target.
Nevertheless, it is challenging and involves massive computational efforts because of the enormous space of
synthetically feasible chemicals \citep{ragoza2022chemsci} and freedom degree of both compound and
protein structures \citep{hawkins2017conformation}.

Recent advances in modeling geometric structures of biomolecules \citep{bronstein2021geometric,atz2021geometric} motivate a promising direction for SBDD \cite{gaudelet2021utilizing,zhang2023systematic}. Several new generative methods have been proposed for the SBDD task~\cite{li2021structure,luo20213d,peng2022pocket2mol,powers2022fragment,ragoza2022generating,zhang2023molecule}, which learn to generate ligand molecules by modeling the complex spatial and chemical interaction features of the binding site. 
For instance, some methods adopt autoregressive models (ARMs) \citep{luo2021autoregressive,liu2022graphbp,peng2022pocket2mol} and show promising results in SBDD tasks, which generate 3D molecules by iteratively adding atoms or bonds based on the target binding site. However, ARMs tend to suffer from error accumulation, and it also is difficult to find an optimal generation order.

\begin{figure}[!tp]
\centering
\includegraphics[width=0.9\linewidth]{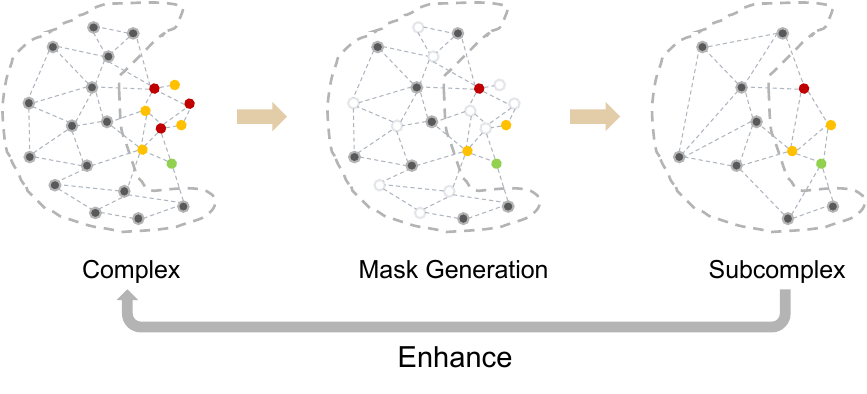}
\caption{\method extracts \textit{subcomplex} from protein-ligand complex, and utilizes it to enhance the binding-adaptive 3D molecule generation in complex.}
\label{fig:motivation}
\end{figure}

An alternative to address these limitations of ARMs is to sample the atomic coordinates and types of
all the atoms at once \citep{du2022molgensurvey}. Recent diffusion-based SBDD methods \citep{guan2023target,schneuing2022structure,lin2022diffbp,guan2023decompdiff} adopt diffusion models \citep{ho2020denoising,song2020score}  to model the distribution of atom types and positions from a standard Gaussian prior with post-processing to assign bonds. 
These diffusion-based methods learn the joint generative process with a SE(3)-equivariant diffusion models \cite{hoogeboom2022equivariant} to capture both spatial and chemical interactions between atoms, and have achieved comparable performance with previous autoregressive models.

Despite the state-of-the-art performance,
existing methods pay little attention to the binding-specific substructure of protein-ligand complex, \textit{i.e.}, the essential part of binding
sites responsible for protein-ligand interactions, which  plays a crucial role in generating molecules with high binding affinities towards the protein targets \cite{bajusz2021exploring,kozakov2015ligand}.
Although recent FLAG \cite{zhang2023molecule} and DrugGPS \cite{zhang2023learning} learn to generate pocket-aware 3D molecules fragment-by-fragment, these massive pre-defined fragments (\textit{e.g.}, motifs or subpockets) are still complex for the model to exactly discover essential protein-ligand interactions from highly diverse protein pockets in nature \cite{spitzer2011surface,basanta2020enumerative}.
Consequently, these limit their practical use in designing high-affinity molecules for new protein targets.

To address these issues, we propose \method, a new binding-adaptive diffusion model for SBDD.
Instead of using pre-defined fragments of pockets or molecules (\textit{e.g.}, subpockets or motifs), at each time step of denoising process, we directly extract essential binding \textit{subcomplex} from protein-ligand complex with a learnable structural pooling. Then we process the selected subcomplex with SE(3)-equivariant GNNs, and transmit them back to the complex as enhanced binding context to improve the atomic target-aware 3D molecule generation.
To facilitate the exchange between the complex and its subcomplex, we iterate the above process via our designed \textit{cross-hierarchy interaction nodes}. Extensive experiments demonstrate that \method can generate molecules with more realistic 3D structures
and higher binding affinities towards the protein targets, while maintaining proper
molecular properties. We highlight our
main contributions as follows:
\begin{itemize}
    \item We propose a hierarchical complex-subcomplex diffusion model for structure-based drug design, which incorporates essential binding-adaptive subcomplex for 3D molecule diffusion generation.
    \item We design and incorporate cross-hierarchy interaction nodes into our iterative denoising networks in the generation process for sufficiently fusing context information.
    \item Empirical results on CrossDocked2020 dataset demonstrate that our \method achieves
    better performance compared with previous methods, higher
    affinity with target protein and other drug properties.
\end{itemize}
\section{Related Work}
\subsection{Structure-Based Drug Design} As the increasing availability of 3D-structure protein-ligand data \cite{kinnings2011machine}, structure-based drug design (SBDD) becomes a hot research area and it aims to generate diverse molecules with high binding affinity to specific protein targets \cite{luo20213d, yang2022knowledge,schneuing2022structure,tan2022target}. Early attempts learn to generate SMILES strings or 2D molecular graphs given protein contexts \cite{skalic2019shape,xu2021novo}. However, it is uncertain whether the resulting compounds with generated strings or graphs could really fit the geometric landscape of the 3D structural pockets. More works start to involve 3D structures of both proteins and molecules
\cite{li2021structure,ragoza2022generating,zhang2023molecule}. \citet{luo20213d}, \citet{liu2022generating}, and \citet{peng2022pocket2mol} adopt autoregressive models to generate 3D molecules in an atom-wise manner.
Recently, powerful diffusion models \cite{sohl2015deep,song2019generative,ho2020denoising} begin to play a role in SBDD, and have achieved promising generation results with non-autoregressive sampling \cite{lin2022diffbp,schneuing2022structure,guan2023target}. TargetDiff \cite{guan2023target}, DiffBP \cite{lin2022diffbp}, and DiffSBDD \cite{schneider1999scaffold} utilize E(n)-equivariant GNNs \cite{satorras2021n} to parameterize conditional diffusion models for protein-aware 3D molecular generation. 
Despite progress, existing methods pay little attention to binding-specific protein-ligand substructures. In contrast, We propose \method to automatically extracts essential binding-adaptive \textit{subcomplex}, and design a hierarchical equivariant molecular diffusion model for SBDD.

\subsection{Diffusion Models for SBDD}
As a new family of deep generative models, diffusion models \citep{sohl2015deep,ho2020denoising,song2020score,yang2022diffusion,yang2023diffusion,yang2023improving} have been recently applied in SBDD tasks. They usually represent the protein-ligand complex by treating
protein binding pockets and ligand molecules as atom point sets in the 3D space, and define a diffusion process for both continuous atom
coordinates and discrete atom types for reverse diffusion generation.
TargetDiff \cite{guan2023target} and DiffBP \cite{lin2022diffbp} both propose a target-aware molecular diffusion process with a SE(3)-equivariant GNN denoiser. 
DecompDiff \cite{guan2023decompdiff} proposes a two-stage diffusion model, which uses an open-source software to obtain molecule-agnostic binding priors as templates for the generation process. In contrast, our \method is a single-stage approach that generates molecules from scratch without relying on external knowledge. It adaptively mines binding-related subcomplexes from the original complex to enhance the generation process, fully considering the interaction between protein pockets and ligands.
\begin{figure*}[ht]
\centering
\includegraphics[width=1.\linewidth]{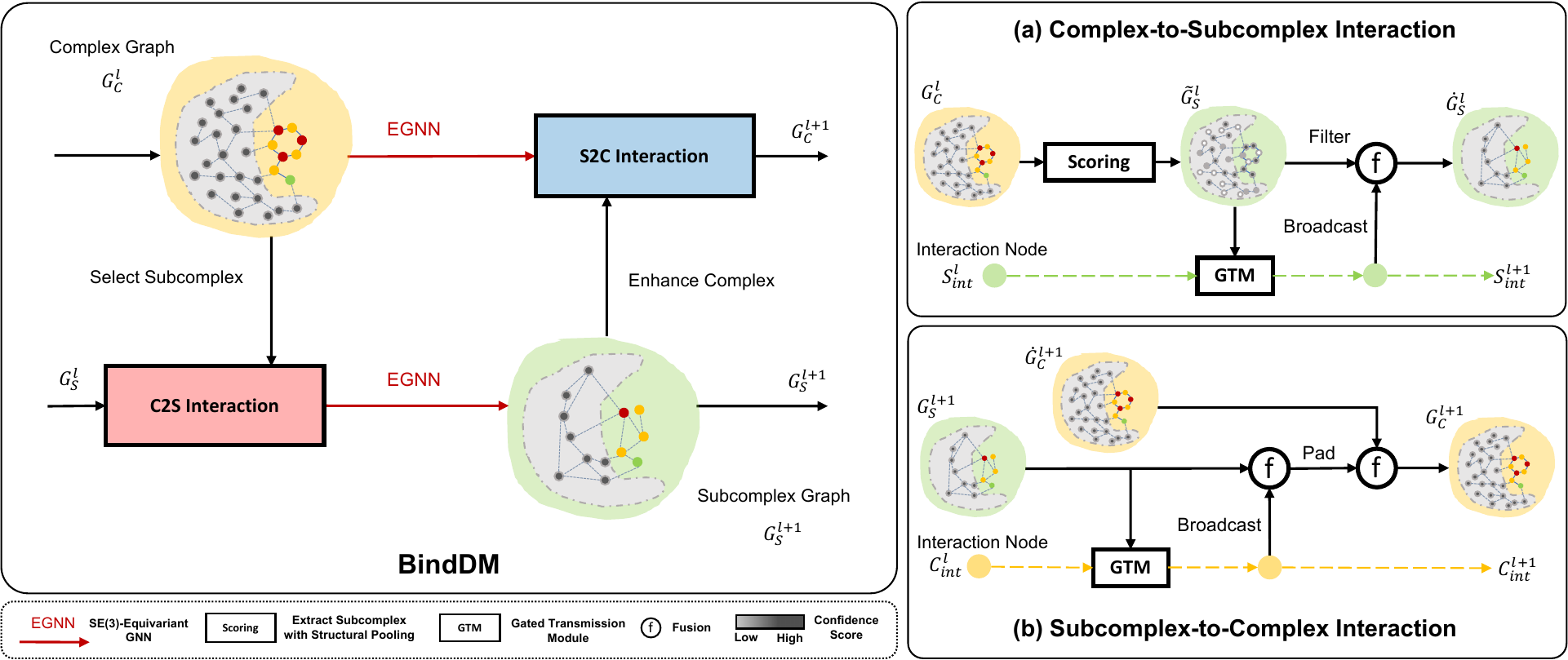}
\caption{The overview of \method.}
\label{fig:subdiff}
\end{figure*}

\section{Preliminary}
\label{sec-preliminary}
The SBDD task from the perspective of generative models can be defined as generating molecules which can bind to a given protein pocket. The target protein and molecule can be represented as $\gP = \{ (\vx_{P}^{(i)}, \vv_{P}^{(i)}) \}_{i=1}^{N_P}$ and $\gM=\{(\vx_{M}^{(i)}, \vv_{M}^{(i)}) \}_{i=1}^{N_M}$, respectively. Here $N_P$ (resp. $N_M$) refers to the number of atoms of the protein $\mathcal{P}$ (resp. the molecule $\mathcal{M}$). $\vx \in \R^3$ and $\vv \in \R^K$ denote the position and type of the atom, respectively. And $K$ denotes the number of atom types.
In the sequel, matrices are denoted by uppercase boldface. For a matrix $\rmX$, $\rvx_i$ denotes the vector on its $i$-th row, and $\rmX_{1:N}$ denotes the submatrix comprising its $1$-st to $N$-th rows. For brevity, the molecule is denoted as $\rmM=[\rmX_M, \rmV_M]$ where $\rmX_M\in \R^{N_M\times 3}$ and $\rmV_M\in \R^{N_M\times K}$, and the protein is denoted as $\rmP=[\rmX_P, \rmV_P]$ where $\rmX_P\in \R^{N_P\times 3}$ and $\rmV_P\in \R^{N_P\times K}$. The task can be formulated as modeling the conditional distribution $p(\rmM|\rmP)$.

Denoising Diffusion Probabilistic Models (DDPMs) equipped with SE(3)-invariant prior and SE(3)-equivariant transition kernel have been applied on the SBDD task \cite{guan2023target,schneuing2022structure,lin2022diffbp}. Specifically, types and positions of the ligand molecule are modeled by DDPM, while the number of atoms $N_M$ is usually sampled from an empirical distribution \cite{hoogeboom2022equivariant,guan2023target} or predicted by a neural network \cite{lin2022diffbp}, and bonds are determined as post-processing.

In the forward diffusion process, a small Gaussian noise is gradually injected into data as a Markov chain. 
Because noises are only added on ligand molecules but not proteins in the diffusion process, we denote the atom positions and types of the ligand molecule at time step $t$ as $\rmX_t$ and $\rmV_t$ and omit the subscript $M$ without ambiguity.  
The diffusion transition kernel can be defined as follows:
\begin{align}\small
\begin{split}
    q(\rmM_t|\rmM_{t-1}, \rmP)=
    &\prod_{i=1}^{N_M}\mathcal{N}(\rvx_{i,t};\sqrt{1-\beta_t}\rvx_{i,t-1},\beta_t\mI)\cdot \\
    &\mathcal{C}(\rvv_{i,t}|(1-\beta_t)\rvv_{i,t-1}+\beta_t/K),
\end{split}
\end{align}
where $\mathcal{N}$ and $\mathcal{C}$ stand for the Gaussian and categorical distribution respectively, $\beta_t$ is defined by fixed variance schedules. The corresponding posterior can be derived as follows: 
\begin{align}\small
\begin{split}
    q(\rmM_{t-1}|\rmM_t,\rmM_0, \rmP)
    =&\prod_{i=1}^{N_M}\mathcal{N}(\rvx_{i,t-1};\Tilde{\vmu}(\rvx_{i,t}, \rvx_{i,0}),\Tilde{\beta}_t\mI) \cdot \\
    &\mathcal{C}(\rvv_{i,t-1}|\Tilde{\vc}(\rvv_{i,t},\rvv_{i,0})),
\end{split}
\end{align}
where 
$\Tilde{\vmu}(\rvx_{i,t},\rvx_{i,0})=\frac{\sqrt{\Bar{\alpha}_{t-1}}\beta_t}{1-\Bar{\alpha}_t}\rvx_{i,0}+\frac{\sqrt{\alpha}_t(1-\Bar{\alpha}_{t-1})}{1-\Bar{\alpha}_t}\rvx_{i,t}$, $\Tilde{\beta}_t=\frac{1-\Bar{\alpha}_{t-1}}{1-\Bar{\alpha}_t}\beta_t$, 
$\alpha_t=1-\beta_t$, 
$\Bar{\alpha}_t=\prod_{s=1}^t \alpha_s$,  
$\Tilde{\vc}(\rvv_{i,t},\rvv_{i,0})=\frac{\vc^*}{\sum_{k=1}^Kc^*_k}$, and $\vc^*(\rvv_{i,t},\rvv_{i,0})=[\alpha_t\rvv_{i,t}+(1-\alpha_t)/K]\odot[\Bar{\alpha}_{t-1}\rvv_{i,0}+(1-\Bar{\alpha}_{t-1})/K]$.

In the approximated reverse process, also known as the generative process, a neural network parameterized by $\theta$ learns to recover data by iteratively denoising. The reverse transition kernel can be approximated with predicted atom types $\hat{\rvv}_{i,0}$ and atom positions $\hat{\rvx}_{i,0}$ as follows:
\begin{align}\small
\begin{split}
    p_\theta(\rmM_{t-1}|\rmM_t, \rmP) = 
    &\prod_{i=1}^{N_M} \mathcal{N}(\rvx_{i,t-1};\Tilde{\vmu}(\rvx_{i,t}, \hat{\rvx}_{i,0}),\Tilde{\beta}_t\mI)\cdot \\
    &\mathcal{C}(\rvv_{i,t-1}|\Tilde{\vc}(\rvv_{i,t},\hat{\rvv}_{i,0})).
\end{split}
\end{align}
\section{The Proposed \method}
As discussed in previous sections, we aim to develop a hierarchical binding-specific diffusion model for SBDD. We here present our proposed \method, as illustrated in \cref{fig:subdiff}.
In this subsection, we will describe how to introduce the selected protein-ligand binding \textit{subcomplex} into the design of the neural network $\phi_\theta$ which predicts (\textit{i.e.}, reconstructs) $\rmM_0=[\rmX_0, \rmV_0]$ in the reverse generation process:
\begin{align}
    [\hat{\rmX}_0, \hat{\rmV}_0] = \phi_\theta([\rmX_t, \rmV_t], t, \rmP).
\end{align}

To extract essential interaction binding-adaptive protein-ligand subcomplex, 
we design a learnable structural pooling to filter the nodes in the original complex graph. To sufficiently utilize both the complex and the subcomplex, we apply SE(3)-equivariant neural networks on them. Finally, we design cross-hierarchy interaction nodes to iteratively exchange information between the complex and the subcomplex, and facilitate the target-aware 3D molecule generation.

\subsection{Binding-Adaptive Subcomplex Extraction}
We first elaborate on how we adaptively extract essential binding subcomplex at each time step $t$. Different from the denoising networks in previous SBDD methods \cite{guan2023target,guan2023decompdiff,schneuing2022structure} that only process the full-atom complex graph, our \method produces a binding-adaptive subcomplex graph from the complex graph with a learnable structural pooling. 
Formally, we have a $k$-nearest neighbors graph $\mathcal{G}_C^{l}$ based on the protein-ligand complex $\rmC=[\rmM_t,\rmP]$ at each denoising time step, where the superscripts $l$ denotes the $l$-th graph layer of the denoising network, and $[\cdot]$ denotes the concatenation along the first dimension.
We aim to extract a binding subcomplex graph $\Tilde{\mathcal{G}}_S^{l}=(\Tilde{\rmH}_S^{l}, \Tilde{\rmX}_S^{l})$ from $\mathcal{G}_C^{l}=(\rmH_C^{l}, \rmX_C^{l})$ (and $\mathcal{G}_S^{l}=(\rmH_S^{l}, \rmX_S^{l})$, for $l>0$), 
where $\rmH \in \R^{N\times d}$ is the node hidden state matrix (initialized with $\rmV_t$ in first layer), $\rmX \in \R^{N\times 3}$ is the node position matrix.
We calculate the confidence scores $\rmZ^{l} \in \R^{N\times 1}$ of all nodes in the complex graph $\mathcal{G}_C^{l}$ contributing to the molecule generation with provided binding sites:
\begin{align}
\hat\rmH^{l}_{C} =&\left\{
\begin{aligned}
    & f_\theta(\rmH^0_{C}), l=0\\
    & f_\theta([\rmH^l_{C}, \text{pad}(\rmH^{l}_{S}, \text{idx}^{l})]), l>0
\end{aligned}
\right. \\
\rmZ^{l} =& \sigma({\rmD^{l}_C}^{-\frac{1}{2}}\rmA^{l}_C {\rmD^{l}_C}^{-\frac{1}{2}}\hat{\rmH}^l_C \Phi_{att}) 
\end{align}
where $\rmA^l_{C} \in \R^{N\times N}$ is the adjacency matrix with pair-wise node connections defined on $k$-nn graph according to $\rmX^l_C$, $\rmD^l_{C} \in \R^{N\times N}$ is the degree matrix of $\rmA^l_{C}$, $f_\theta(\cdot)$ is an MLP, $\Phi_{att}$ is the learnable parameter, $\text{pad}(\cdot)$ is the operation of filling empty nodes into the position of filtered nodes according to the indices of selected nodes. In this way, the padded subcomplex graph has the same number of nodes as the complex graph. The $\text{idx}^l$ is indices of the top $\lceil rN \rceil$ nodes which are selected based on confidence scores $\rmZ^l$, and $r\in (0, 1]$ is the selection ratio that determines the number of nodes to keep:
\begin{align}
    \text{idx}^l = \text{top-rank}(\rmZ^l, \lceil rN \rceil)
\end{align}
where the top-rank is the function that returns the indices of the top $\lceil rN \rceil$ values. In practice, we set $r=0.5$.
Then, the hidden state matrix $\Tilde{\rmH}_S^{l}$ and position matrix $\Tilde{\rmX}_S^{l}$ of subcomplex are obtained: 
\begin{align}
    \Tilde{\rmH}_S^{l} =& \rmH_{C, \text{idx}^l, :}^{l} \odot \rmZ^l_{\text{idx}^l}, \\
    \Tilde{\rmX}_S^{l} =& \rmX_{C, \text{idx}^l, :}^{l}
\end{align}
where $\cdot_{\text{idx}^l}$ is an indexing operation, $\odot$ is the broadcasted elementwise product, $\rmH_{C, \text{idx}^l, :}^l$ and $\rmX_{C, \text{idx}^l, :}^l$ are the row-wise (i.e. node-wise) indexed  matrix $\rmH_{C}^l$ and $\rmX_{C}^l$, respectively. Next, we process the selected subcomplex to better leverage binding context for molecule generation.

\subsection{3D Equivariant Complex-Subcomplex Processing} 
Our goal is to generate 3D molecules based on target protein binding sites, the model needs to
generate both continuous atom coordinates and discrete atom types, while being SE(3)-equivariant to global translation and rotation during the entire generative process. 
This property is a critical inductive bias for generating 3D molecules \cite{hoogeboom2022equivariant,schneuing2022structure,guan2023target}, and an invariant distribution composed with an equivariant transition function will result in an invariant distribution.
Thus, for our hierarchical complex-subcomplex denoising network, we have the following proposition in the setting of protein-aware molecule generation.
\begin{proposition}
Denoting SE(3)-transformation as $T$, we can achieve invariant likelihood
\textit{w.r.t} $T$ on both the protein-ligand complex and its subcomplex: $p_\theta(T\rmM_0|T\rmP) = p_\theta(\rmM_0|\rmP)$ if we shift the Center of Mass (CoM) of protein atoms to zero and parameterize the Markov transition $p(\rvx_{t-1}|\rvx_t,\rvx_P)$ with a
SE(3)-equivariant network.
\end{proposition}
We apply two SE(3)-equivariant neural networks on the $k$-nn graphs ($\mathcal{G}_C^{l}$ and $\dot{\mathcal{G}}_S^{l}$) of the protein-ligand complex and its corresponding subcomplex in the denoising process, respectively. 
For the subcomplex graph $\dot{\mathcal{G}}_S^{l}$ updated through the complex-to-subcomplex (C2S) interaction, the SE(3)-invariant hidden states $\dot{\rmH}_S^l$ and SE(3)-equivariant positions $\dot{\rmX}_S^l$ are updated as follows to obtain the ${\mathcal{G}}_S^{l+1}$: 
\begin{align}\small
\begin{split}
    \rvh_{S,i}^{l+1}=& \dot{\rvh}_{S,i}^l+\sum_{j\in\mathcal{N}_{i}}f^l_{S,h}\left(\left\Vert\dot{\rvx}^l_{S,i}-\dot{\rvx}^l_{S,j}\right\Vert,\dot{\rvh}_{S,i}^l,\dot{\rvh}_{S,j}^l,\rve_{S,ij}\right) \\
    \rvx_{S,i}^{l+1}= &
    \dot{\rvx}_{S,i}^l +
    \sum_{j\in\mathcal{N}_{i}}
    \left(\dot{\rvx}_{S,i}^l-\dot{\rvx}_{S,j}^l\right) \cdot \\
    &f_{S,x}^l
    \left(\left\Vert\dot{\rvx}^l_{S,i}-\dot{\rvx}^l_{S,j}\right\Vert,\rvh_{S,i}^{l+1},\rvh_{S,j}^{l+1},\rve_{S,ij}\right)\cdot\mathds{1}_{\text{mol}}
\end{split}
\end{align}
where $\mathcal{N}_{i}$ is the set of $k$-nearest neighbors of atom $i$ on the subcomplex graph, $\rve_{S,ij}$ indicates the atom $i$ and atom $j$ are both protein atoms or both ligand atoms or one protein atom and one ligand atom, and $\mathds{1}_{\text{mol}}$ is the ligand atom mask since the protein atom coordinates are known and thus supposed to remain unchanged during this update. 
The similar process are applied on the complex graph $\mathcal{G}_C^{l} = (\rmH_C^l, \rmX_C^l)$ to obtained the $\dot{\mathcal{G}}_C^{l+1} = (\dot{\rmH}_C^{l+1}, \dot{\rmX}_C^{l+1})$.

\begin{figure}[ht]
\centering
\includegraphics[width=0.75\linewidth]{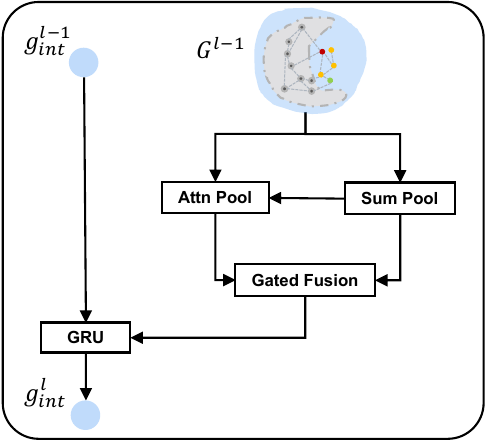}
\caption{The overview of the gated transmission module.}
\label{fig:GTM}
\end{figure}

\subsection{Iterative Cross-Hierarchy Interaction}
In \method, we introduce two \textit{cross-hierarchy interaction nodes} to facilitate the information exchange between the binding contexts of two hierarchies, the complex graph and its subcomplex graph.
Specifically, we initialize the interaction node of subcomplex graph via sum pooling:
\begin{align}
    \rvc_{int}^{l} = \text{Pool}_{\text{sum}}(\mathcal{G}^{l}_S).
\end{align}
Then the binding context of the extracted subcomplex graph is transmitted back to the complex graph $\mathcal{G}^{l}_C$ for cross-hierarchy information fusion via the interaction node $\rvc_{int}^l$ and the gated transmission module as shown in \cref{fig:GTM}:
\begin{align}
& \hat{\rvc}_{int}^{l} = \text{Attn}(\text{Query}(\rvc_{int}^{l}), \text{Key}(\rmH_{S}^{l}))\cdot \text{Value}(\rmH_{S}^{l}) \label{eq-1}\\
& \alpha_{\rvc}^{l} = \sigma(f_{{1}}({\rvc}_{int}^{l}) + f_{{2}}(\hat{{\rvc}}_{int}^{l})) \label{eq-2}\\
& {\rvc}_{int}^{l} = \text{GRU}({\rvc}_{int}^{l-1}, \alpha_{\rvc}^{l} \cdot {\rvc}_{int}^{l} + (1-\alpha_{\rvc}^{l}) \cdot \hat{{\rvc}}_{int}^{l}) \label{eq-3}\\
& \rvh_{C,i}^{l} = f_3(\dot{\rvh}_{C,i}^{l}, {\rvc}_{int}^{l}) \label{eq-4}
\end{align}
where $\sigma$ is sigmoid, $f_{1}$, $f_{2}$ and $f_3$ are MLPs, $\dot{\rvh}_{C,i}^{l}$ is the SE(3)-invariant hidden state of $i$-th node in $\dot{\mathcal{G}}_C^{l}$. 
\cref{eq-1,eq-2,eq-3} mix the messages between the subcomplex graph and the interaction node through gated recurrent unit (GRU) \cite{chung2014empirical} for updating $\rvc_{int}^{l}$. \cref{eq-4} perform node-wise fusion with $\rvc_{int}^{l}$ for subcomplex-to-complex (S2C) interaction. Similarly, we also have the interaction node $\rvs_{int}^{l}$ for complex-to-subcomplex (C2S) interaction, and we iterate these cross-hierarchy processes for sufficiently incorporating binding-adaptive subcomplex into the 3D molecule generation process as shown in \cref{fig:subdiff}.

\subsection{Training and Sampling}
To train \method (\textit{i.e.}, optimize the evidence lower bound induced by \method), we use the same objective function as \citet{guan2023target}.
The atom position loss and atom type loss at time step $t-1$ are defined as follows respectively:
\begin{align}
\begin{split}
    \mathbf{L}^{(x)}_{t-1}
    & =\frac{1}{2\Tilde{\beta}^2_t} 
    \sum_{i=1}^{N_M} 
    \Vert \Tilde{\vmu}(\rvx_{i,t},\rvx_{i,0}) - \Tilde{\vmu}(\rvx_{i,t}, \hat{\rvx}_{i,0}) \Vert^2 \\
    & =\gamma_t \sum_{i=1}^{N_M} \Vert \rvx_{i,0} - \hat{\rvx}_{i,0} \Vert \label{eq:l_x}\\
\end{split}
\end{align}
\begin{align}
    \mathbf{L}^{(v)}_{t-1} 
    =\sum_{i=1}^{N_M}\sum_{k=1}^K \Tilde{\vc}(\rvv_{i,t}, \rvv_{i,0})_k \log \frac{\Tilde{\vc}(\rvv_{i,t}, \rvv_{i,0})_k}{\Tilde{\vc}(\rvv_{i,t}, \hat{\rvv}_{i,0})_k}\label{eq:l_v}
\end{align}
where 
$\hat{\rmX}_{0}$ and $\hat{\rmV}_{0}$ 
are predicted from $\rmX_t$ and $\rmV_t$, and $\gamma_t=\frac{\bar{\alpha}_{t-1}\beta_t^2}{2\Tilde{\beta}_t^2(1-\Bar{\alpha}_t)^2}$. Kindly recall that $\rvx_{i,t}$,  $\rvv_{i,t}$, $\hat{\rvx}_{i,0}$, and $\hat{\rvv}_{i,0}$ correspond to the $i$-th row of $\rmX_t$, $\rmV_t$, $\hat{\rmX}_{0}$, and $\hat{\rmV}_{0}$, respectively. 
The final loss is a weighted sum of atom coordinate loss and atom type loss with a hyperparameter $\lambda$ as: $\mathbf{L}=\mathbf{L}^{(x)}_{t-1}+\lambda \mathbf{L}^{(v)}_{t-1}$.
\section{Experiments}
\subsection{Experimental Settings}
\label{sec-experimental_settings}
\paragraph{Datasets and Baseline Methods} 
As for molecular generation, following the previous work~\cite{luo20213d, peng2022pocket2mol,guan2023target}, we train and evaluate \method on the CrossDocked2020 dataset~\cite{francoeur2020three}. Adhering to the data preparation and splitting procedures outlined by~\citet{luo20213d}, we refine the $22.5$ million docked binding complexes to high-quality docking poses (RMSD between the docked pose and the ground truth $<1$\AA) and diverse proteins (sequence identity $<$ $30\%$). This meticulous process produces $100,000$ protein-ligand pairs for training and $100$ proteins for testing.
We compare our model against four recent representative methods for SBDD. \textbf{LiGAN} \cite{ragoza2022chemsci} is a conditional VAE model trained on an atomic density grid representation of protein-ligand structures. \textbf{AR} \cite{luo20213d} and \textbf{Pocket2Mol} \cite{peng2022pocket2mol} are autoregressive schemes that generate 3D molecules atoms conditioned on the protein pocket and previous generated atoms. \textbf{TargetDiff} \cite{guan2023target} and \textbf{DecompDiff} \cite{guan2023decompdiff} represent state-of-the-art diffusion methods, generating atom coordinates and atom types in a non-autoregressive manner.

\paragraph{Evaluation}
We conduct a comprehensive assessment of the generated molecules, evaluating them from three key perspectives: \textbf{molecular structures}, \textbf{target binding affinity}, and \textbf{molecular properties}. In terms of \textbf{molecular structures}, we quantify the Jensen-Shannon divergences (JSD) in empirical distributions of atom/bond distances between the generated molecules and the reference ones.
To evaluate the \textbf{target binding affinity}, following previous work \cite{luo20213d,ragoza2022generating,guan2023target}, we adopt AutoDock Vina~\citep{eberhardt2021autodock} to compute and report the mean and median of binding-related metrics, including \textit{Vina Score}, \textit{Vina Min}, \textit{Vina Dock} and \textit{High Affinity}. Vina Score directly estimates the binding affinity; Vina Min involves local structure minimization before estimation; Vina Dock integrates an additional re-docking process for optimal binding affinity; High affinity measures the ratio of generated molecules binding better than the reference molecule per test protein.
To evaluate \textbf{molecular properties}, we utilize \textit{QED}, \textit{SA}, \textit{Diversity} as metrics following~\citet{luo20213d, ragoza2022chemsci}. QED is a quantitative estimation of drug-likeness combining several desirable molecular properties; SA (synthesize accessibility) estimates the difficulty of synthesizing the ligands; Diversity is computed as average pairwise dissimilarity between all generated ligands. 
All sampling and evaluation procedures follow~\citet{guan2023target} for fair comparison.  

\begin{table}[ht]
    \centering
    \begin{adjustbox}{width=0.4\textwidth}
    \renewcommand{\arraystretch}{1.2}
    \begin{tabular}{c|c|c|c|c|c|c}
    \hline
    {Bond} & liGAN & AR & \thead{Pocket2\\Mol} & \thead{Target\\Diff} & \thead{Decomp\\Diff} & Ours \\
    \hline
    C$-$C & 0.601 & 0.609 & 0.496 & 0.369 & \textbf{0.359} & \underline{0.380} \\ 
    C$=$C & 0.665 & 0.620 & 0.561 & 0.505 & 0.537 & \textbf{0.229} \\ 
    C$-$N & 0.634 & 0.474 & 0.416 & 0.363 & 0.344 & \textbf{0.265} \\ 
    C$=$N & 0.749 & 0.635 & 0.629 & 0.550 & 0.584 & \textbf{0.245} \\ 
    C$-$O & 0.656 & 0.492 & 0.454 & 0.421 & 0.376 & \textbf{0.329} \\ 
    C$=$O & 0.661 & 0.558 & 0.516 & 0.461 & 0.374 & \textbf{0.249} \\ 
    C$:$C & 0.497 & 0.451 & 0.416 & 0.263 & \textbf{0.251} & \underline{0.282}  \\ 
    C$:$N & 0.638 & 0.552 & 0.487 & 0.235 & 0.269 & \textbf{0.130} \\
    \hline
    \end{tabular}
    \renewcommand{\arraystretch}{1}
    \end{adjustbox}
    \caption{Jensen-Shannon divergence between bond distance distributions of reference and generated molecules, lower values indicate better performances. ``-'', ``='', and ``:'' represent single, double, and aromatic bonds, respectively.
    }
    \label{tab:jsd}
\end{table}
\begin{table*}[ht]
    \centering
    \begin{adjustbox}{width=0.9\textwidth}
    \renewcommand{\arraystretch}{1.2}
    \begin{tabular}{l|cc|cc|cc|cc|cc|cc|cc}
    \hline
    \multirow{2}{*}{Methods} & \multicolumn{2}{c|}{Vina Score ($\downarrow$)} & \multicolumn{2}{c|}{Vina Min ($\downarrow$)} & \multicolumn{2}{c|}{Vina Dock ($\downarrow$)} & \multicolumn{2}{c|}{High Affinity ($\uparrow$)} & \multicolumn{2}{c|}{QED ($\uparrow$)}   & \multicolumn{2}{c|}{SA ($\uparrow$)} & \multicolumn{2}{c}{Diversity ($\uparrow$)} \\
     & Avg. & Med. & Avg. & Med. & Avg. & Med. & Avg. & Med. & Avg. & Med. & Avg. & Med. & Avg. & Med. \\
    \hline
    Reference   & -6.36 & -6.46 & -6.71 & -6.49 & -7.45 & -7.26 & -  & - & 0.48 & 0.47 & 0.73 & 0.74 & - & -   \\
    \hline
    LiGAN       & - & - & - & - & -6.33 & -6.20 & 21.1\% & 11.1\% & 0.39 & 0.39 & 0.59 & 0.57 & 0.66 & 0.67  \\
    
    GraphBP     & - & - & - & - & -4.80 & -4.70 & 14.2\% & 6.7\% & 0.43 & 0.45 & 0.49 & 0.48 & \textbf{0.79} & \textbf{0.78}   \\
    
    AR          & \underline{-5.75} & -5.64 & -6.18 & -5.88 & -6.75 & -6.62 & 37.9\% & 31.0\% & \underline{0.51} & 0.50 & \underline{0.63} & \underline{0.63} & 0.70 & 0.70  \\
    
    Pocket2Mol  & -5.14 & -4.70 & -6.42 & -5.82 & -7.15 & -6.79 & 48.4\% & 51.0\% & \textbf{0.56} & \textbf{0.57} & \textbf{0.74} & \textbf{0.75} & 0.69 & 0.71  \\
    
    TargetDiff  & -5.47 & \underline{-6.30} & -6.64 & -6.83 & -7.80 & -7.91 & 58.1\% & 59.1\% & 0.48 & 0.48 & 0.58 & 0.58 & 0.72 & 0.71  \\

    DecompDiff  & -5.67 & -6.04 & \underline{-7.04} & \underline{-6.91} & \underline{-8.39} & \textbf{-8.43} & \underline{64.4}\% & \underline{71.0}\% & 0.45 & 0.43 & 0.61 & 0.60 & 0.68 & 0.68  \\
    
    \method     & \textbf{-5.92} & \textbf{-6.81} & \textbf{-7.29} & \textbf{-7.34} & \textbf{-8.41} & \underline{-8.37} & \textbf{64.8}\% & \textbf{71.6}\% & \underline{0.51} & \underline{0.52} & 0.58 & 0.58 & \underline{0.75} & \underline{0.74}  \\
    
    \hline
    \end{tabular}
    \renewcommand{\arraystretch}{1}
    \end{adjustbox}
    \caption{Summary of different properties of reference molecules and molecules generated by our model and other baselines. ($\uparrow$) / ($\downarrow$) denotes a larger / smaller number is better. Top 2 results are highlighted with \textbf{bold text} and \underline{underlined text}, respectively. 
    }
    \label{tab:main_tab}
\end{table*}
\subsection{Main Results}
\paragraph{Generated 3D Molecular Structures}
We compare our \method and the representative methods in terms of molecular structures.
We compute different bond distributions of the generated molecules and compare them against the corresponding reference empirical distributions in~\cref{tab:jsd}.
Our model has a comparable performance with TargetDiff and DecompDiff and substantially outperforms all other baselines across all major bond types, indicating the great potential of \method for generating stable molecular structures.

\begin{figure*}[ht]
\centering
\includegraphics[width=0.9\linewidth]{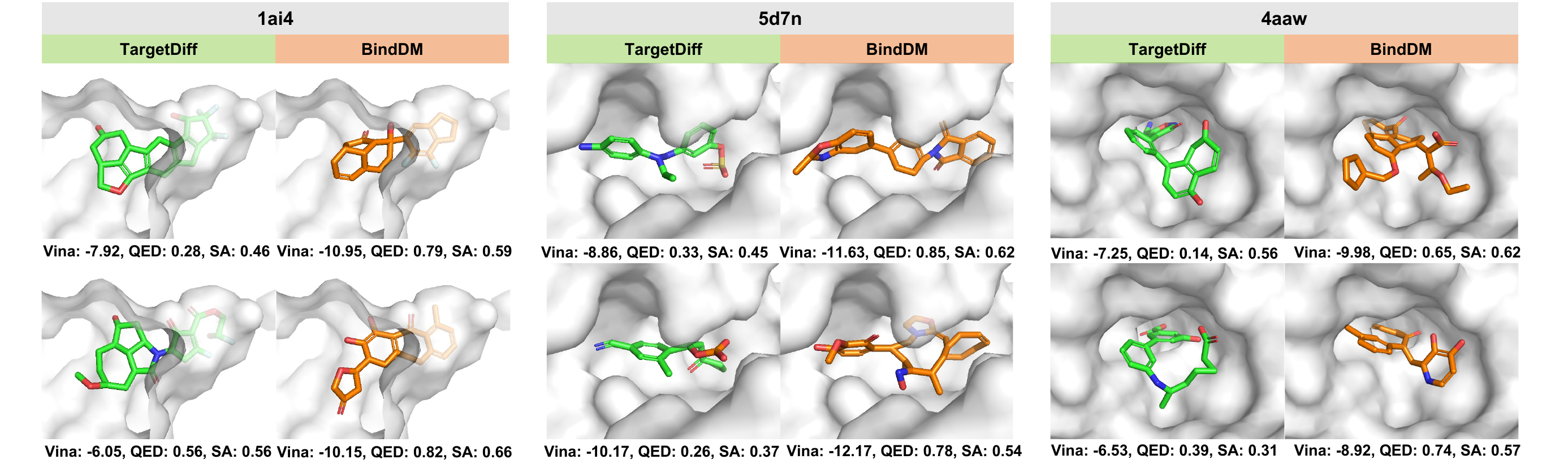}
\caption{The generated ligand molecules of TargetDiff~\citep{guan2023target} and \method for the given protein pockets. Carbon atoms in ligands generated by TargetDiff and \method are visualized in green and orange, respectively. We report Vina Score, QED, SA for each molecule.}
\label{fig:viz}
\end{figure*}

\paragraph{Target Binding Affinity and Molecule Properties}
We evaluate the effectiveness of \method in terms of binding affinity. 
We can see in \cref{tab:main_tab} that our \method outperforms baselines in binding-related metrics.
Specifically, \method surpasses strong autoregressive method Pocket2Mol by a large margin of \textbf{15.2\%}, \textbf{13.6\%} and \textbf{17.6\%} in Avg. Vina Score, Vina Min and Vina Dock, respectively. 
Compared with the state-of-the-art diffusion-based method DecompDiff, \method not only increased the binding-related metrics Avg. Vina Score and Vina Min by \textbf{4.4\%} and \textbf{3.6\%}, respectively, but also significantly increased the property-related metric Avg. QED by \textbf{13.3\%}.
In terms of high-affinity binder, we find that on average \textbf{64.8\%} of the \method molecules show better binding affinity than the reference molecule, which is significantly better than other baselines. 
These gains demonstrate that the proposed \method effectively captures significant binding-related subcomplex to enable generating molecules with improved target binding affinity.
Moreover, we can see a trade-off between property-related metrics QED and binding-related metrics in previous methods. 
DecompDiff performs better than AR and Pocket2Mol in binding-related metrics, but falls behind them in QED scores. In contrast, our \method not only achieves the state-of-the-art binding-related scores but also maintains proper QED score, achieving a better trade-off than DecompDiff.
Nevertheless, we put less emphasis on QED and SA because they are often applied as rough screening metrics in real drug discovery scenarios, and it would be fine as long as they are within a reasonable range. 
\cref{fig:viz} shows some examples of generated ligand molecules and their properties. The molecules generated by our model have valid structures and reasonable binding poses to the target, which are supposed to be promising candidate ligands.

\subsection{Model Analysis}
\paragraph{Effect of the Iterative Cross-Hierarchy Interaction on Target-specific Molecule Generation}
We conduct a set of ablation experiments to study the effect of iterative cross-hierarchy interaction on the generation ability of diffusion models for the target-specific molecules: 
(1) \textbf{Exp0}: the baseline model without applying the iterative cross-hierarchy interaction, 
(2) \textbf{Exp1}: we replace the binding-adaptive subcomplex extraction (BASE) module with a random selection of atoms from the complex for constructing the subcomplex. The selection ratio is set to 0.5,
(3) \textbf{Exp2}: we remove subcomplex graphs in iterative cross-hierarchy interaction and keep interaction nodes $\rvc_{int}$ unchanged for information propagation between cross-layer complex graphs, 
(4) \textbf{Exp3}: we remove interaction nodes $\rvc_{int}$ and $\rvs_{int}$ and keep the extracted subcomplex graphs unchanged, 
(5) \textbf{Exp4}: we remove the gated transmission module in the update of interaction nodes $\rvc_{int}$ and $\rvs_{int}$.
The results are present in \cref{tab:abla_component}.

In the comparison between Exp0 and Exp1, we can find that randomly selected subcomplex can not provide useful information about pocket-ligand binding. 
And the comparison between Exp1 and \method suggests that BASE is more effective than random selection in exploring binding-related clues from the complex. The effectiveness of BASE is beneficial for \method in generating molecules that are tightly bound to the given protein pocket. 
In comparing Exp2 with \method, it is evident that solely relying on global interaction nodes for information propagation between cross-layer complex graphs does not provide significant binding-related information for pocket-specific molecular generation.
In comparing Exp3 with \method, we observe that the utilization of global interaction nodes for information exchange between complex and subcomplex not only improves the performance of \method in binding-related metrics but also contributes to the molecular property-related ones. 
And the same conclusion is also observed in the comparison between Exp4 and \method.

\begin{table}[ht]
    \centering
    \begin{adjustbox}{width=0.45\textwidth}
    \renewcommand{\arraystretch}{1.2}
    \begin{tabular}{l|cc|cc|cc|cc}
    \hline
    \multirow{2}{*}{Methods} & \multicolumn{2}{c|}{Vina Score ($\downarrow$)} & \multicolumn{2}{c|}{Vina Min ($\downarrow$)} & \multicolumn{2}{c|}{Vina Dock ($\downarrow$)} & \multicolumn{2}{c}{QED ($\uparrow$)} \\
     & Avg. & Med. & Avg. & Med. & Avg. & Med. & Avg. & Med. \\
    \hline
    Exp0   & -5.04 & -5.75 & -6.38 & -6.52 & -7.55 & -7.72 & 0.46 & 0.46  \\
    
    Exp1   & -4.79 & -5.92 & -6.36 & -6.66 & -7.71 & -7.63 & \underline{0.50} & \underline{0.51}  \\
    
    Exp2   & \underline{-5.65} & -6.25 & -6.64 & -6.65 & -7.96 & -7.77 & 0.45 & 0.45 \\    

    Exp3   & -5.62 & \underline{-6.74} & \underline{-6.83} & \underline{-6.92} & \underline{-8.11} & \underline{-8.15} & 0.47 & 0.46  \\

    Exp4   & -5.60 & -6.28 & -6.78 & -6.83 & -7.94 & -8.01 & 0.47 & 0.47  \\

    \method     & \textbf{-5.92} & \textbf{-6.81} & \textbf{-7.29} & \textbf{-7.34} & \textbf{-8.41} & \textbf{-8.37} & \textbf{0.51} & \textbf{0.52} \\
    
    \hline
    \end{tabular}
    \renewcommand{\arraystretch}{1}
    \end{adjustbox}
    \caption{Effect of the iterative cross-hierarchy interaction on target-specific molecule generation. ($\uparrow$) / ($\downarrow$) denotes a larger / smaller number is better. Top 2 results are highlighted with \textbf{bold text} and \underline{underlined text}, respectively.
    }
    \label{tab:abla_component}
\end{table}

\paragraph{Influence of Extracting Binding Clues from Complex, Pocket and Ligand}
Since the presence of binding clues in both the molecular ligands and protein pockets, we conduct three experiments to explore the effects of extracting binding-related clues from different structures on how tightly the generated molecules bind to the specific protein pockets:
the binding-related substructures are extracted from (1) the molecular ligands, (2) the protein pockets, and (3) the complexes (treating the molecule and pocket as a unified entity) to enhance the generation of molecular ligands binding tightly to specific protein pockets, respectively.
As present in \cref{tab:abla_anchor_type}, \method can achieve the best performance on binding-related metrics when binding-related substructures are extracted from complexes and used to enhance the generation process of protein-specific molecular ligands.

\begin{table}[!h]
    \centering
    \begin{adjustbox}{width=0.45\textwidth}
    \renewcommand{\arraystretch}{1.2}
    \begin{tabular}{l|cc|cc|cc|cc}
    \hline
    \multirow{2}{*}{Methods} & \multicolumn{2}{c|}{Vina Score ($\downarrow$)} & \multicolumn{2}{c|}{Vina Min ($\downarrow$)} & \multicolumn{2}{c|}{Vina Dock ($\downarrow$)} & \multicolumn{2}{c}{QED ($\uparrow$)} \\
     & Avg. & Med. & Avg. & Med. & Avg. & Med. & Avg. & Med. \\
    \hline
    baseline     & -5.04 & -5.75 & -6.38 & -6.52 & -7.55 & -7.72 & \underline{0.46} & \underline{0.46}  \\

    Pocket   & -5.37 & \textbf{-6.84} & \underline{-7.03} & \textbf{-7.38} & \underline{-8.30} & \underline{-8.36} & \textbf{0.51} & \textbf{0.52}   \\

    Ligand   & \underline{-5.46} & -6.77 & -6.98 & -7.27 & -8.13 & 8.29 & \textbf{0.51} & \textbf{0.52}  \\

    Complex     & \textbf{-5.92} & \underline{-6.81} & \textbf{-7.29} & \underline{-7.34} & \textbf{-8.41} & \textbf{-8.37} & \textbf{0.51} & \textbf{0.52}  \\
    
    \hline
    \end{tabular}
    \renewcommand{\arraystretch}{1}
    \end{adjustbox}
    \caption{Influence of extracting binding clues from complex, pocket and ligand. ($\uparrow$) / ($\downarrow$) denotes a larger / smaller number is better. Top 2 results are highlighted with \textbf{bold text} and \underline{underlined text}, respectively.
    }
    \label{tab:abla_anchor_type}
\end{table}

\begin{figure}[ht]
\centering
\includegraphics[width=0.7\linewidth]{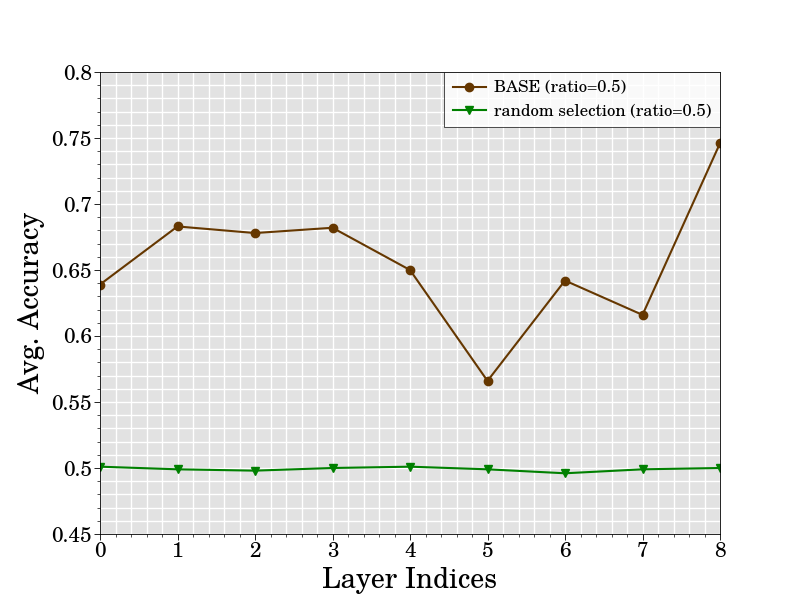}
\caption{The subcomplex prediction accuracy of BASE in each layer of the de-nosing network.}
\label{fig:subcomp_acc}
\end{figure}

\paragraph{Correlation between Adaptively Extracted Subcomplex and Pocket-Ligand Binding Clues}
\label{para:binding_clues}
To validate the presence of binding-related clues in the subcomplex extracted by BASE in \method and their suitability for generating molecular ligands tightly bound to specific protein pockets, we initially employed the pre-trained binding affinity prediction model BAPNet~\cite{li2021structure}.  
By predicting binding affinity of the complex which consists of the given protein pocket and the molecules generated by \method, the binding-related subcomplex is obtained by ranking the complex atoms according to contributions to the binding affinity prediction.
We take the subcomplexes predicted by BAPNet as the reference subcomplexes, and calculate the accuracy of the subcomplexes predicted by BASE as a metric to assess the effectiveness of BASE in extracting binding-related subcomplex.
Considering that the sampling process consists of 1000 steps, we calculate the average accuracy by comparing the reference subcomplex with all the subcomplexes extracted by each layer of the denoising network in \method (9 layers in total) throughout the entire sampling process.
As shown in \cref{fig:subcomp_acc}, each layer of the denoising network of \method achieves an accuracy rate of around 0.65 for subcomplex extracted through BASE, which is higher than the accuracy rate of around 0.5 for random selection (in practice, the selection ratio is set to 0.5). 
This suggests that BASE in \method is capable of extracting binding-related subcomplex to a certain degree.
In addition, we replace the process of using BASE to adaptively select binding-related subcomplex with randomly selecting atoms from the complex to construct a substructure. The performance present in~\cref{tab:abla_component} (Exp1) demonstrates the subcomplex extracted from BASE benefit the final performance.
 
\section{Conclusion}
In this paper, we propose an effective diffusion model \method to adaptively extract the essential part of binding sites responsible for protein-ligand interactions, \textit{subcomplex}, for enhancing protein-aware 3D molecule generation. We further design \textit{cross-hierarchy interaction node} to facilitate the hierarchical information exchange between complex and subcomplex. 
Empirical results sufficiently demonstrate \method can generate more realistic 3D molecules with higher binding affinities towards the protein targets, while maintaining proper molecular properties. 
For future work, we will extend our \method to few-shot scenarios \citep{yang2020dpgn} by leveraging recent advances in graph representation learning \citep{yang2023vqgraph,yang2022omni}.
\section{Acknowledgement}
The work was partly supported by the National Natural Science Foundation of China (No.62171251), the Major Key Project of GZL under Grant SRPG22-001 and the Major Key Project of PCL under Grant PCL2023A09.

\bibliography{aaai24}

\appendix
\section{Training and Sampling Procedure of \method}
Following TargetDiff \cite{guan2023target}, the training and sampling procedure of \method as summarized as follows:

\begin{algorithm}[h]
\caption{Training Procedure of \method}\label{appendix:training}
\textbf{Input}: Protein-ligand binding dataset $\{\gP, \gM\}_{i=1}^N$ as described in \textbf{Preliminary}, denoising network $\phi_\theta$\\
\textbf{Output}: Denoising network $\phi_\theta$
\begin{algorithmic}[1] 
\WHILE{$\phi_\theta$ not converge}
\STATE Sample diffusion time $t \in \gU(0, \dots, T)$
\STATE Move the complex to make CoM of protein atoms zero
\STATE Perturb $[\rmX_0,\rmV_0]$ to obtain $[\rmX_t,\rmV_t]$
\STATE Embed $\rmV_t$ into $\rmH^0_M$, and embed $\rmV_P$ into $\rmH^0_P$
\STATE Predict $[\hat\rmX_0, \hat\rmV_0]$ from $[\rmX_t, \rmH_M]$ and $[\rmX_P, \rmH_P]$ 
\STATE Compute loss $\mathbf{L}$ with $[\hat\rmX_0, \hat\rmV_0]$ and $[\rmX_M, \rmV_M]$
\STATE Update $\theta$ by minimizing $\mathbf{L}$
\ENDWHILE
\end{algorithmic}
\end{algorithm}

\begin{algorithm}[h]
\caption{Sampling Procedure of \method}\label{appendix:sampling}
\textbf{Input}: The protein binding site $\gP$, the learned denoising network $\phi_\theta$\\
\textbf{Output}: Generated ligand molecule $\gM$ that binds to the protein pocket $\gP$
\begin{algorithmic}[1] 
\STATE Sample the number of atoms $N_M$ of the ligand molecule $\gM$ as described in \textbf{Preliminary}
\STATE Move CoM of protein atoms to zero
\STATE Sample initial ligand atom coordinates $\rvx_T$ and atom types $\rvv_T$
\FOR{$t$ in $T, T-1, \dots, 1$}
\STATE Embed $\rmV_t$ into $\rmH_M$
\STATE Predict $[\hat\rmX_0, \hat\rmV_0]$ from $[\rmX_t, \rmH_M]$ and $[\rmX_P, \rmH_P]$
\STATE Sample $\rmX_{t-1}$ and $\rmV_{t-1}$ from the posterior $p_\theta(\rmM_{t-1}|\rmM_t, \rmP)$ 
\ENDFOR
\end{algorithmic}
\end{algorithm}

\section{Details of Hierarchical Complex-Subcomplex Denoising Network in \method}
\paragraph{Input Initialization}
Following TargetDiff \cite{guan2023target}, we use a one-hot element indicator \{H, C, N, O, S, Se\} and one-hot amino acid type indicator (20 types) to represent each protein atom. Similarly, we represent each ligand atom using a one-hot element indicator \{C, N, O, F, P, S, Cl\}. Additionally, we introduce a one-dimensional flag to indicate whether the atoms belong to the protein or ligand. Two 1-layer MLPs are introduced to map the inputs of protein and ligand into 128-dim spaces respectively.
For representing the connection between atoms, we introduce a 4-dim one-hot vector to indicate four bond types: bond between protein atoms, ligand atoms, protein-ligand atoms or ligand-protein atoms. And we introduce distance embeddings by using the distance with radial basis functions located at 20 centers between 0 Å and 10 Å. Finally we calculate the outer products of distance embedding and bond types to obtain the edge features.

\paragraph{Model Architectures}
At the $l$-th layer, we dynamically construct the protein-ligand complex and subcomplex with a $k$-nearest neighbors (knn) graph based on coordinates of the given protein and the ligand from previous layer. In practice, we set the number of neighbors $k_n=32$. And we apply an SE(3)-equivariant neural network for message passing. The 9-layer equivariant neural network consists of Transformer layers with 128-dim hidden layer and 16 attention heads. Following \citet{guan2023target}, in the diffusion process, we select the fixed sigmoid $\beta$ schedule with $\beta_1=1\mathrm{e}{-7}$ and $\beta_T=2\mathrm{e}{-3}$ as variance schedule for atom coordinates, and the cosine $\beta$ schedule with $s=0.01$ for atom types. The number of diffusion steps are set to 1000.  

\paragraph{Training Details}
We use the Adam as our optimizer with learning rate 0.001, $betas=(0.95, 0.999)$, batch size 4 and clipped gradient norm 8.  We balance the atom type loss and atom position loss by multiplying a scaling factor $\lambda=100$ on the atom type loss.

\section{More Ablation Studies}
\paragraph{Time Complexity}
For investigating the sampling efficiency, we report the inference time of our model and other baselines for generating 100 valid molecules on average. Pocket2Mol, TargetDiff and DecompDiff use 2037s, 1987s and 3218s, and \method takes 2851s / 3372s when the selection ratio $r$ are set to 0.3 and 0.5 respectively. It worth noting that, DecompDiff is a two-stage method which needs to obtain the priors through the external software, and TargetDiff and \method are single-stage methods.
Besides, the inference time of \method with different selection ratio $r$ are present in the \cref{appendix:select_ratio}.

\paragraph{Effect of Subcomplex Selection Ratio}
We conduct a series of experiments to explore the impact of different selection ratios in BASE on the performance of \method. As present in \cref{appendix:select_ratio}, when the selection ratio is set to 0.1, \method adds almost no computational complexity compared to the baseline, yet it still achieves a significant improvement in performance. Notably, \method achieves the best performance on both binding- and molecular property-related metrics when the selection ratio is set to 0.5.

\begin{table*}[htbp]
    \centering
    \begin{adjustbox}{width=1.0\textwidth}
    \renewcommand{\arraystretch}{1.2}
    \begin{tabular}{l|cc|cc|cc|cc|cc|cc|cc|c}
    \hline
    \multirow{2}{*}{Methods} & \multicolumn{2}{c|}{Vina Score ($\downarrow$)} & \multicolumn{2}{c|}{Vina Min ($\downarrow$)} & \multicolumn{2}{c|}{Vina Dock ($\downarrow$)} & \multicolumn{2}{c|}{High Affinity ($\uparrow$)} & \multicolumn{2}{c|}{QED ($\uparrow$)}   & \multicolumn{2}{c|}{SA ($\uparrow$)} & \multicolumn{2}{c|}{Diversity ($\uparrow$)} & Inference Time ($\downarrow$) \\
     & Avg. & Med. & Avg. & Med. & Avg. & Med. & Avg. & Med. & Avg. & Med. & Avg. & Med. & Avg. & Med. & (100 molecules) \\
    \hline
    baseline     & -5.04 & -5.75 & -6.38 & -6.52 & -7.55 & -7.72 & 54.2\% & 54.1\% & 0.46 & 0.46 & 0.57 & 0.57 & 0.71 & 0.69 & \textbf{1729} s  \\

    r=0.1   & -5.61 & -6.58 & -7.00 & -7.12 & -8.19 & -8.23 & 63.4\% & 62.2\% & 0.50 & 0.51 & \textbf{0.59} & \textbf{0.59} & \underline{0.74} & \underline{0.73} & \underline{2060} s \\

    r=0.3   & -5.64 & -6.64 & \underline{-7.11} & \underline{-7.22} & \underline{-8.35} & \underline{-8.33} & \textbf{66.0}\% & \underline{67.3}\% & \textbf{0.52} & \textbf{0.53} & \underline{0.58} & \underline{0.58} & \underline{0.74} & \textbf{0.74} & 2851 s \\

    r=0.5   & \textbf{-5.92} & \textbf{-6.81} & \textbf{-7.29} & \textbf{-7.34} & \textbf{-8.41} & \textbf{-8.37} & 64.8\% & \textbf{71.6}\% & \underline{0.51} & \underline{0.52} & \underline{0.58} & \underline{0.58} & \textbf{0.75} & \textbf{0.74} & 3372 s \\

    r=0.7   & \underline{-5.74} & \underline{-6.64} & \underline{-7.08} & -7.18 & -8.22 & -8.17 & \underline{64.7}\% & 66.2\% & \underline{0.51} & \underline{0.52} & \textbf{0.59} & \textbf{0.59} & \underline{0.74} & \textbf{0.74} & 3553 s \\

    r=0.9   & -5.20 & -6.48 & -6.97 & -7.12 & -8.20 & -8.31 & 62.5\% & 66.3\% & 0.50 & \underline{0.52} & \underline{0.58} & \underline{0.58} & \underline{0.74} & \underline{0.73} & 3795 s  \\
    
    \hline
    \end{tabular}
    \renewcommand{\arraystretch}{1}
    \end{adjustbox}
    \caption{Effect of different subcomplex selection ratios $r$.
    }
    \label{appendix:select_ratio}
\end{table*}

\paragraph{Complete Results}
The complete results (including all evaluation metrics) of \cref{tab:abla_component} and \cref{tab:abla_anchor_type} are present in \cref{appendix:abla_component} and \cref{appendix:abla_anchor_type}, respectively.
\begin{table*}[htbp]
    \centering
    \begin{adjustbox}{width=1.0\textwidth}
    \renewcommand{\arraystretch}{1.2}
    \begin{tabular}{l|cc|cc|cc|cc|cc|cc|cc}
    \hline
    \multirow{2}{*}{Methods} & \multicolumn{2}{c|}{Vina Score ($\downarrow$)} & \multicolumn{2}{c|}{Vina Min ($\downarrow$)} & \multicolumn{2}{c|}{Vina Dock ($\downarrow$)} & \multicolumn{2}{c|}{High Affinity ($\uparrow$)} & \multicolumn{2}{c|}{QED ($\uparrow$)}   & \multicolumn{2}{c|}{SA ($\uparrow$)} & \multicolumn{2}{c}{Diversity ($\uparrow$)} \\
     & Avg. & Med. & Avg. & Med. & Avg. & Med. & Avg. & Med. & Avg. & Med. & Avg. & Med. & Avg. & Med. \\
    \hline
    Exp0   & -5.04 & -5.75 & -6.38 & -6.52 & -7.55 & -7.72 & 54.2\% & 54.1\% & 0.46 & 0.46 & 0.57 & 0.57 & 0.71 & 0.69  \\
    
    Exp1   & -4.79 & -5.92 & -6.36 & -6.66 & -7.71 & -7.63 & 57.9\% & 53.4\% & \underline{0.50} & \underline{0.51} & \underline{0.59} & \underline{0.58} & 0.72 & 0.70  \\
    
    Exp2   & \underline{-5.65} & -6.25 & -6.64 & -6.65 & -7.96 & -7.77 & 61.6\% & 60.8\% & 0.45 & 0.45 & \textbf{0.60} & \textbf{0.59} & 0.70 & 0.72  \\    

    Exp3   & -5.62 & \underline{-6.74} & \underline{-6.83} & \underline{-6.92} & \underline{-8.11} & \underline{-8.15} & \underline{62.7}\% & \underline{63.3}\% & 0.47 & 0.46 & 0.58 & \textbf{0.59} & \underline{0.74} & 0.73  \\

    Exp4   & -5.60 & -6.28 & -6.78 & -6.83 & -7.94 & -8.01 & 62.2\% & 62.5\% & 0.47 & 0.47 & 0.56 & 0.55 & 0.73 & \textbf{0.75}  \\

    \method     & \textbf{-5.92} & \textbf{-6.81} & \textbf{-7.29} & \textbf{-7.34} & \textbf{-8.41} & \textbf{-8.37} & \textbf{64.8}\% & \textbf{71.6}\% & \textbf{0.51} & \textbf{0.52} & 0.58 & \underline{0.58} & \textbf{0.75} & \underline{0.74}  \\
    
    \hline
    \end{tabular}
    \renewcommand{\arraystretch}{1}
    \end{adjustbox}
    \caption{Effect of the iterative cross-hierarchy interaction on target-specific molecule generation. ($\uparrow$) / ($\downarrow$) denotes a larger / smaller number is better. Top 2 results are highlighted with \textbf{bold text} and \underline{underlined text}, respectively.
    }
    \label{appendix:abla_component}
\end{table*}

\begin{table*}[htbp]
    \centering
    \begin{adjustbox}{width=1.0\textwidth}
    \renewcommand{\arraystretch}{1.2}
    \begin{tabular}{l|cc|cc|cc|cc|cc|cc|cc}
    \hline
    \multirow{2}{*}{Methods} & \multicolumn{2}{c|}{Vina Score ($\downarrow$)} & \multicolumn{2}{c|}{Vina Min ($\downarrow$)} & \multicolumn{2}{c|}{Vina Dock ($\downarrow$)} & \multicolumn{2}{c|}{High Affinity ($\uparrow$)} & \multicolumn{2}{c|}{QED ($\uparrow$)}   & \multicolumn{2}{c|}{SA ($\uparrow$)} & \multicolumn{2}{c}{Diversity ($\uparrow$)} \\
     & Avg. & Med. & Avg. & Med. & Avg. & Med. & Avg. & Med. & Avg. & Med. & Avg. & Med. & Avg. & Med. \\
    \hline
    baseline     & -5.04 & -5.75 & -6.38 & -6.52 & -7.55 & -7.72 & 54.2\% & 54.1\% & \underline{0.46} & \underline{0.46} & \underline{0.57} & \underline{0.57} & 0.71 & 0.69  \\

    Pocket   & -5.37 & \textbf{-6.84} & \underline{-7.03} & \textbf{-7.38} & \underline{-8.30} & \underline{-8.36} & \underline{64.2}\% & \underline{63.5}\% & \textbf{0.51} & \textbf{0.52} & \underline{0.57} & \underline{0.57} & \underline{0.74} & \textbf{0.74}  \\

    Ligand   & \underline{-5.46} & -6.77 & -6.98 & -7.27 & -8.13 & -8.29 & 63.6\% & 62.3\% & \textbf{0.51} & \textbf{0.52} & \textbf{0.58} & \textbf{0.58} & \textbf{0.75} & \textbf{0.74}  \\

    Complex     & \textbf{-5.92} & \underline{-6.81} & \textbf{-7.29} & \underline{-7.34} & \textbf{-8.41} & \textbf{-8.37} & \textbf{64.8}\% & \textbf{71.6}\% & \textbf{0.51} & \textbf{0.52} & \textbf{0.58} & \textbf{0.58} & \textbf{0.75} & \textbf{0.74}  \\
    
    \hline
    \end{tabular}
    \renewcommand{\arraystretch}{1}
    \end{adjustbox}
    \caption{Influence of extracting binding clues from complex, pocket and ligand. ($\uparrow$) / ($\downarrow$) denotes a larger / smaller number is better. Top 2 results are highlighted with \textbf{bold text} and \underline{underlined text}, respectively.
    }
    \label{appendix:abla_anchor_type}
\end{table*}

\section{More Results}
We provide the visualization of more ligand molecules generated by \method, comparing to both reference and TargetDiff \cite{guan2023target}, as shown in \cref{appendix:more_viz}.
\begin{figure*}[htbp]
\centering
\includegraphics[width=1.0\linewidth]{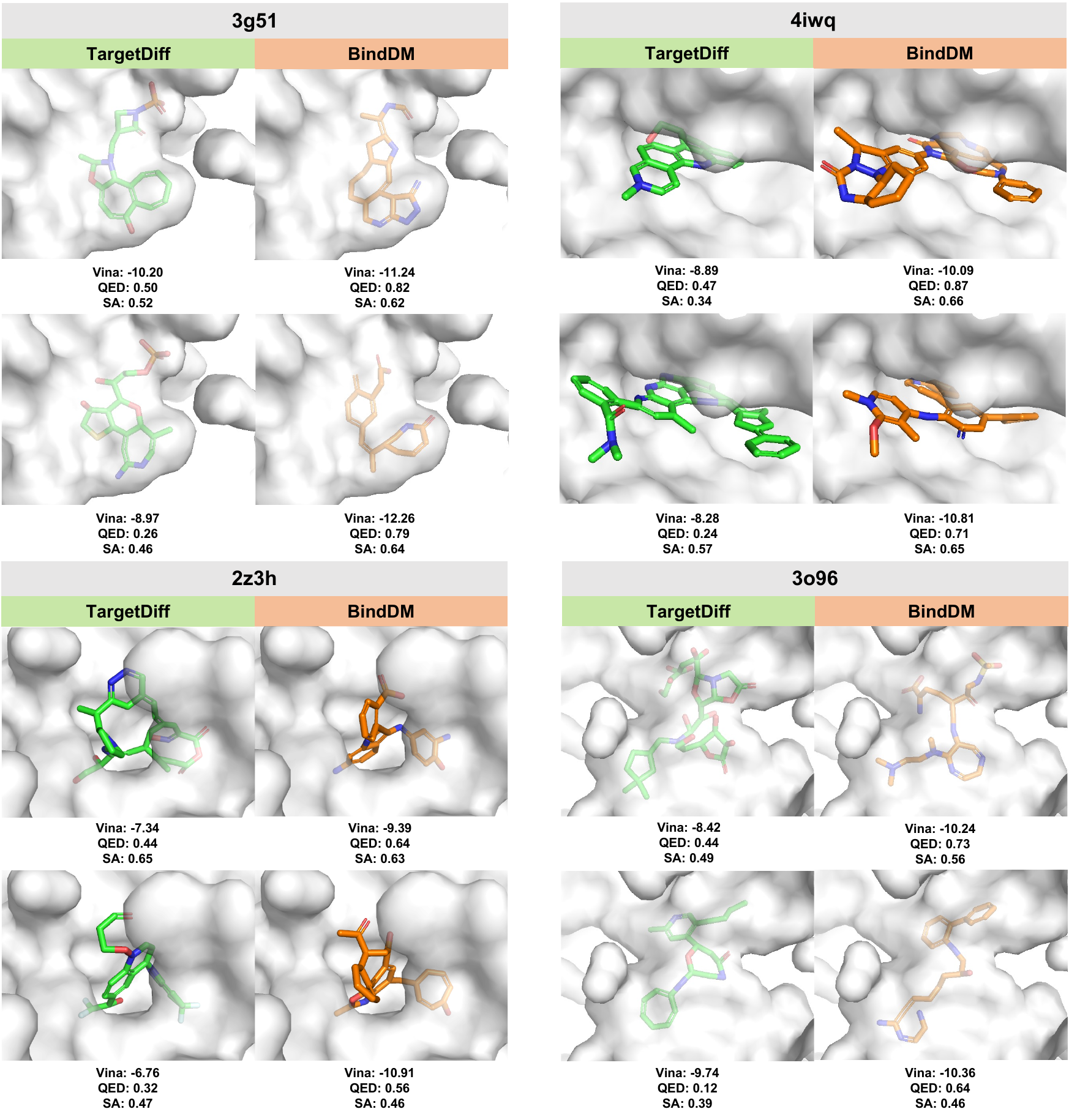}
\caption{The generated ligand molecules of TargetDiff~\citep{guan2023target} and \method for the given protein pockets. Carbon atoms in ligands generated by TargetDiff and \method are visualized in green and orange, respectively. We report Vina Score, QED, SA for each molecule.}
\label{appendix:more_viz}
\end{figure*}


\end{document}